\documentclass[useAMS,usenatbib]{mn2e}

\usepackage {graphicx}
\usepackage{color}

\title[Barium and Yttrium abundance in intermediate-age and old 
open clusters]{Barium and Yttrium abundance in  intermediate-age and old open 
clusters\thanks{Based on observations collected at ESO VLT under programs 076.B-0263 
and 083.D-0682.}}

\author[T.~Mishenina et al.]{T.~Mishenina$^{1}$\thanks{E-mail:tmishenina@ukr.net},
S.~Korotin$^{1}$, G.~Carraro$^{2,3}$, V.V.~Kovtyukh$^{1}$, and I.A.~Yegorova$^{2}$
\\
$^1$Astronomical Observatory of Odessa National University,
and Isaac Newton Institute of Chile Odessa branch,
Odessa, Ukraine\\
$^2$European Southern Observatory, Chile\\
$^3$Dipartimento di Fisica e Astronomia, Universit\'a di Padova, Italy}

\begin{document}

\date{Accepted. Received ; in original form }

\pagerange{\pageref{firstpage}--\pageref{lastpage}} \pubyear{2013}

\maketitle

\label{firstpage}

\begin{abstract}
Barium is a neutron capture element, that, in open clusters, is frequently
over-abundant with respect to the Iron. A clear explanation for this is still 
missing.
Additionally, its gradient across the Galactic disk is poorly constrained.
We measure the abundance of yttrium and barium  using the synthetic spectrum 
method from UVES high-resolution spectra of eight distant open clusters, namely 
Ruprecht 4, Ruprecht 7, Berkeley 25, Berkeley 73, Berkeley 75,
NGC~6192, NGC~6404,  and NGC~6583. The 
barium abundance was estimated using NLTE approximation.
We confirm that Barium is indeed over-abundant in most clusters, 
especially young clusters.
Finally, we investigated the trend of yttrium and barium abundances as a function 
of distance in the Galaxy and ages.
Several scenarios for the barium over-abundance are then  discussed.
\end{abstract}

\begin{keywords}
stars: abundances -- stars: late-type -- Galaxy: disc -- Galaxy: evolution.
\end{keywords}

\section{Introduction}

There is nowadays much interest in the investigation of the abundances of 
elements produced via  neutron capture processes (e.g. Sr, Ba, Y, La, Zr, and so forth),
both in open clusters and in 
the general Galactic field.
The precise abundance of these elements in as many stars and clusters as possible 
allows us to improve our understanding  (1) of the neutron capture processes 
themselves as enrichment sources (e.g. \citealt{bu01}), 
(2) of the various chemical evolution scenarios for the Galactic disk
(e.g. \citealt{chi97}; \citealt{ser09} ),   
and (3) of the different formation paths of star clusters and stars in the 
Galactic field \citep{BC06}.

The neutron capture reactions can run in two ways, depending on the neutron
flux density: as a slow process (s-process) or as a rapid process (r-process, 
\citep{BBFH57}. 
The two processes take place at different temperatures, in different time 
domains, and, therefore, inside objects in different evolutionary stages.

As for the r-process, three channels 
of s-process were originally introduced in the classical model by \citet{kap89} : 
(1) the main channel, from asymptotic giant branch (AGB) stars (isotopes with A 
from 90 to 204); (2) the weak channel,  from massive stars that explode later as
Supernovae (isotopes with A up to 90); and (3) the strong channel, introduced 
to reproduce 50\% of $^{208}$Pb. 
At present, modern models of nucleosynthesis in the AGB stars consider a 
variety of  contributions in the production of specific isotopes, as a function 
of  (1) the number of third dredge-up episodes, and (2) the yields of the light 
s-elements (ls), as  Sr, Y, Zr, and the heavy s-elements (hs), such as Ba, La, 
Ce, Pr, Nd,  and their dependence on metallicity (see more detailed information
in \citealt{bu99}; \citealt{gal98}; \citealt{ser09}, etc.) 
The small and moderately massive AGB stars (mass around 1.5 solar masses) 
contribute mostly to the  production of the n-capture elements, such as yttrium 
and barium . Those stars then enrich the interstellar medium, from which new 
stars will form in subsequent star formation episodes.

 The details of the formation and production of neutron capture elements provide, 
therefore, crucial information bearing on models of Galactic chemical evolution.
For instance, in the study by \citet{dor09}, barium abundances were derived for 
a number of open clusters' stars, and it was found that Ba abundance with 
respect to iron increases at decreasing clusters age. 
This seems to require some additional enrichment with barium 
by the least-massive stars $M < 1.5 \sim M_{\sun}$  (\citealt{bu99}; \citealt{str09}; \citealt{cri09}), and it is at variance with 
the currently existing chemical evolution models that describe the enrichment 
process at the moment of the Solar System formation \citep{tra99}. 
Along the same vein, \citet{jfp11} and \citet{des11} found significant barium 
overabundances in young stars.

\citet{mai11} investigated other n-capture elements, such 
as Y, Zr, La, Ce, in young open clusters and confirmed the result obtained for 
barium,  namely an increase of overabundances of those elements with the 
cluster age. 
However, the overabundances are limited to small values ([El/Fe] $\sim$ 
0.2 dex), when compared with barium overabundance ([Ba/Fe] $\sim$ 0.6 dex). 

In the recent study by \citet{dor12}, a lower value of 0.3 dex of the barium 
abundance was obtained for the stars in three young associations. 

The authors provide a few possible  explanations for  this overabundance, namely : 
(1) neglecting the hyperfine structure of the barium lines; 
(2) deviations from the Local Thermodynamic Equilibrium (LTE) conditions; 
(3) chromospheric activity (see also \citealt{dor09}). 
 However, these  authors did not detect any correlation between the 
barium abundance and the chromospheric activity indices for their stars. 
Moreover, their estimate of the deviations from the LTE conditions implied
a minor effect on the barium abundance determination.  

As is  shown in a number of works (e.g. \citealt{mai11}; \citealt{des11}; 
\citealt{dor12}), the origin of the barium overabundance (up to 0.6 dex) is 
still far from being understood,  and much work still need to  be done.
 
The goal of the present study is  (1) to enlarge the sample of star clusters 
with estimates of LTE yttrium abundances and  the barium abundances 
allowing  for deviations from the LTE, and (2) to analyze the trends of  
their abundances as a function of the distance to the Galactic center and the 
age. This is in an attempt to provide further observational pieces of evidence,
and to boost the discussion on this intriguing problem.

\section{Open clusters: observations and the stellar parameters}

The main parameters (Galactic coordinates, helio-centric distance and age) of 
the investigated clusters are listed in the Table \ref{par}. 

\begin {table}
\caption {The main parameters of the clusters under study}
\label {par}
\begin{tabular}{l c c c c c}
\hline
 Name       &     l   &     b   &$d_{\sun}$ &    age  & Ref.  \\
            &    deg  &    deg  &     kpc   &    Gyr  &       \\
\hline
Berkeley 75 &  234.30&    11.12 &     9.8   &    3.0  &  Ca05a \\
Berkeley 25 &  226.60&     9.69 &    11.3   &    4.0  &  Ca05a \\
Ruprecht 7  &  225.44&     4.58 &     6.5   &    0.8  &  Ca05b \\
Ruprecht 4  &  222.04&     5.31 &     4.9   &    0.8  &  Ca05b \\
Berkeley 73 &  215.28&     9.42 &     9.7   &    1.5  &  Ca05a \\
NGC6192     &  340.65&     2.12 &     1.5   &    0.18 &  Cl    \\
NGC6404     &  355.66&    -1.18 &     1.7   &    0.5  &  Ca05c \\
NGC6583     &    9.28&    -1.53 &     2.1   &    1.0  &  Ca05c \\
\hline  
\end {tabular}  

\medskip
Ca05a stands for \citet{car05a}\\
Ca05b stands for \citet{car05b}\\
Ca05c stands for \citet{car05c}\\
Cl    stands for \citet{cla06} \\
\end {table}

\begin {table}
\caption {Comparison of the estimated effective temperatures. Ca07 stands for Carraro et al. 2007.}
\label {compar}
\begin{tabular}{l c c c c c}
\hline
Cluster	    &  $T_{\rm eff}$ ,K    &  $T_{\rm eff}$, K & sigma & N &  $T_{\rm eff}$ error\\
       	    &  Ca07 &  this study  &       &   &  K     \\
\hline
Ruprecht4\_3 &    5105             & 5112         & 106   & 51& 15 \\  
Ruprecht7\_2 &    5160             & 5195         & 146   & 54& 20 \\ 
\hline  
\end {tabular}  
\end {table}

%

\begin{figure}
\resizebox{\hsize}{!}
{\includegraphics{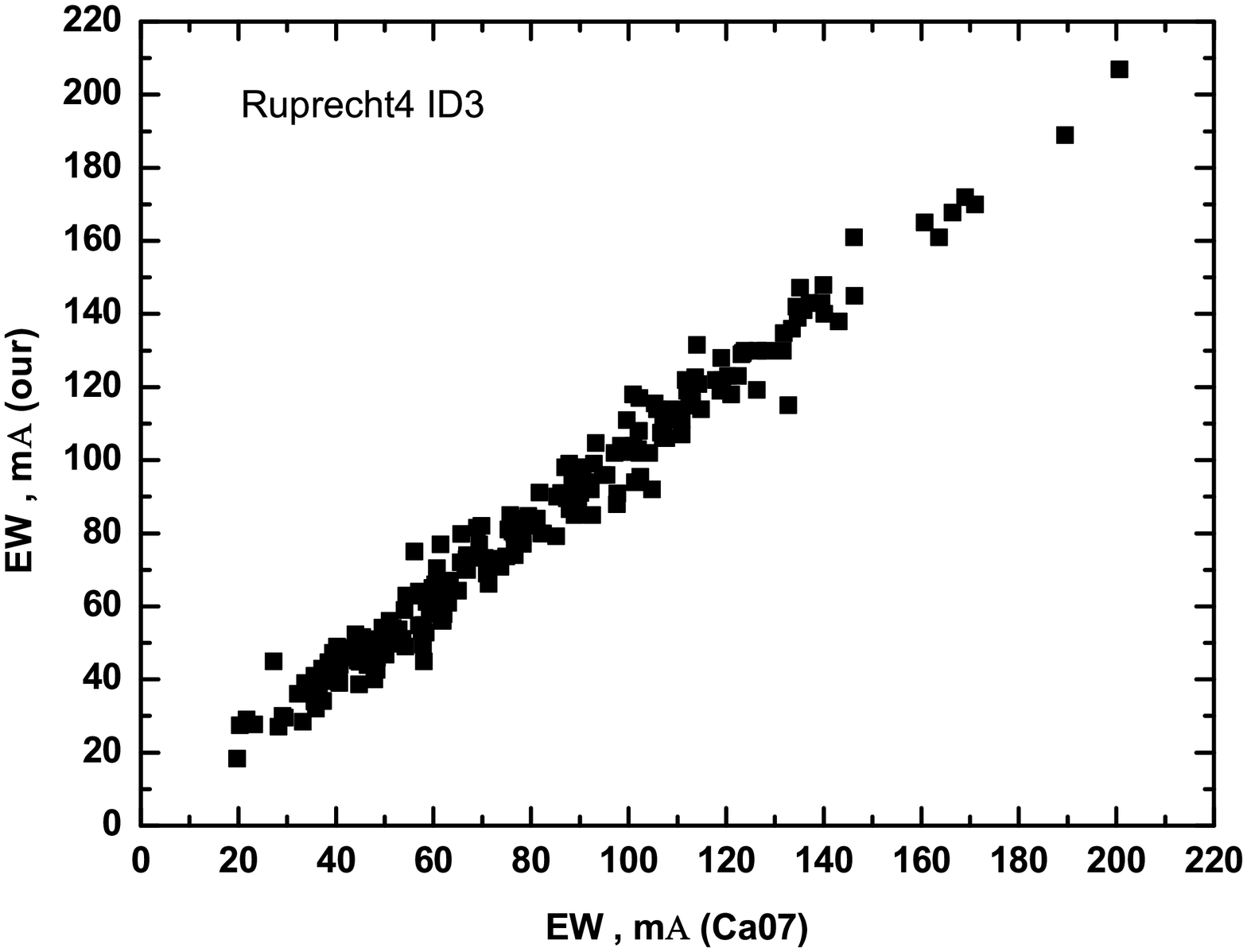}}
\resizebox{\hsize}{!}
{\includegraphics{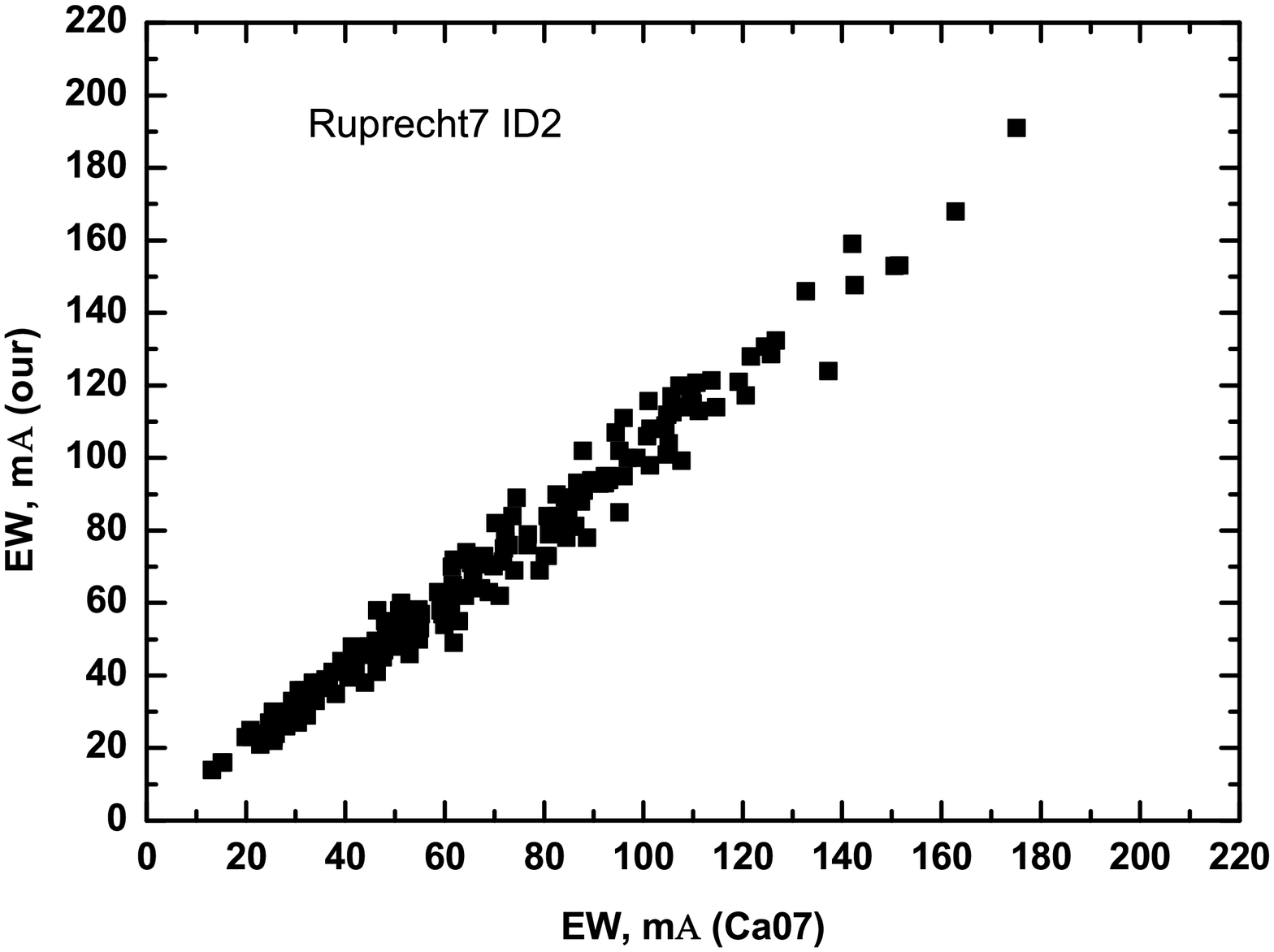}}
\caption[]{The comparison of the equivalent widths for star Ruprecht4\_3
and Ruprecht7\_2.}
\label{ewr47}
\end{figure}

The observational material used in this study was taken from \citet{car07} and 
\citet{mag10}. The spectra were obtained using the high-resolution 
echelle spectrograph UVES onboard VLT with resolving power R = 40,000 for the 
wavelengths range 4750--6800~\AA. 

Additionally, in \citet{car07} radial velocities were measured and the 
membership to the  clusters established. 

In order to corroborate the  atmospheric parameters obtained in \citet{car07},
we selected  two stars in Ruprecht~4 and Ruprecht~7, respectively,  and  
compared the equivalent widths of the lines with the ones measured in the 
present study using the DECH20 software package \citep{gal92}. 
We also compared the temperatures derived by \citet{car07}, with our effective 
temperatures, estimated with the method developed by \citet{kov03}, which 
consists in calibrating the central depths ratio of the lines with different 
potentials for the lower excitation level.
The comparison of the equivalent widths for Ruprecht4\_3 and Ruprecht7\_2, is 
shown in 
Fig. \ref{ewr47}, while the comparison 
of the estimated effective temperatures for the same stars is presented in 
Table \ref{compar}.

Having obtained a good agreement for both the equivalent widths and 
effective temperatures, we are now in a good position to estimate the yttrium 
and barium abundances in the remaining stars by using the atmospheric 
parameters obtained by \citet{car07} and \citet{mag10}.
 In these works the metallicity and element abundances 
were determined using the analysis code MOOG \citep{sn73} and the atmospheric models 
by \citet{kur93}. The effective temperature was estimated from  photometry 
using  the \citet{AAMR99} calibration. Then the temperature was refined using 
the correlation between  iron abundance, determined with that line, and the 
potential of the lower level of that line. The surface gravity was computed 
with the canonical formula (for more details see \citealt{car07}).
                                                                                              
\section{Determination of the yttrium and barium abundances}

\begin{figure*}
\resizebox{\hsize}{!}
{\includegraphics{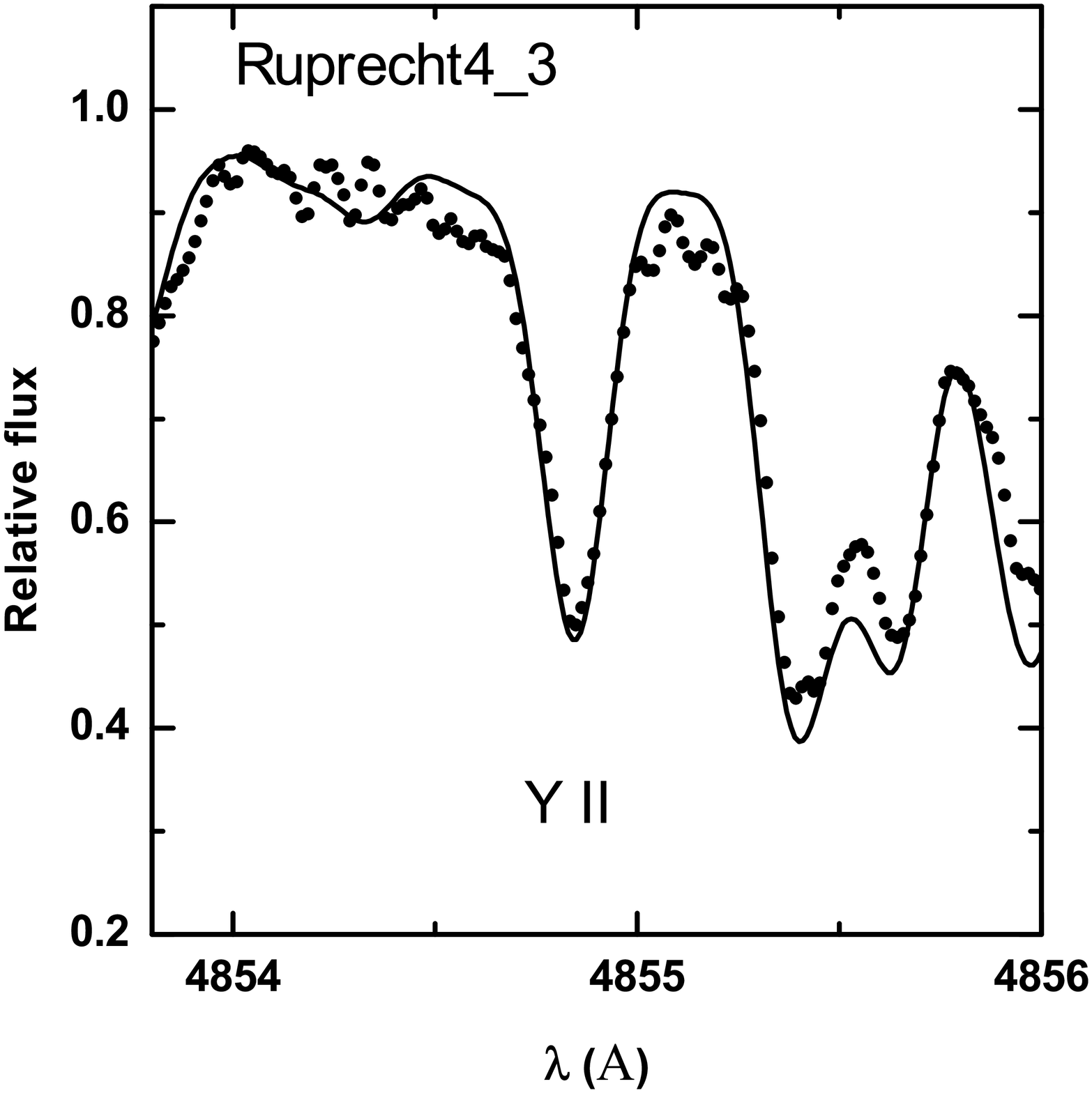}
\includegraphics{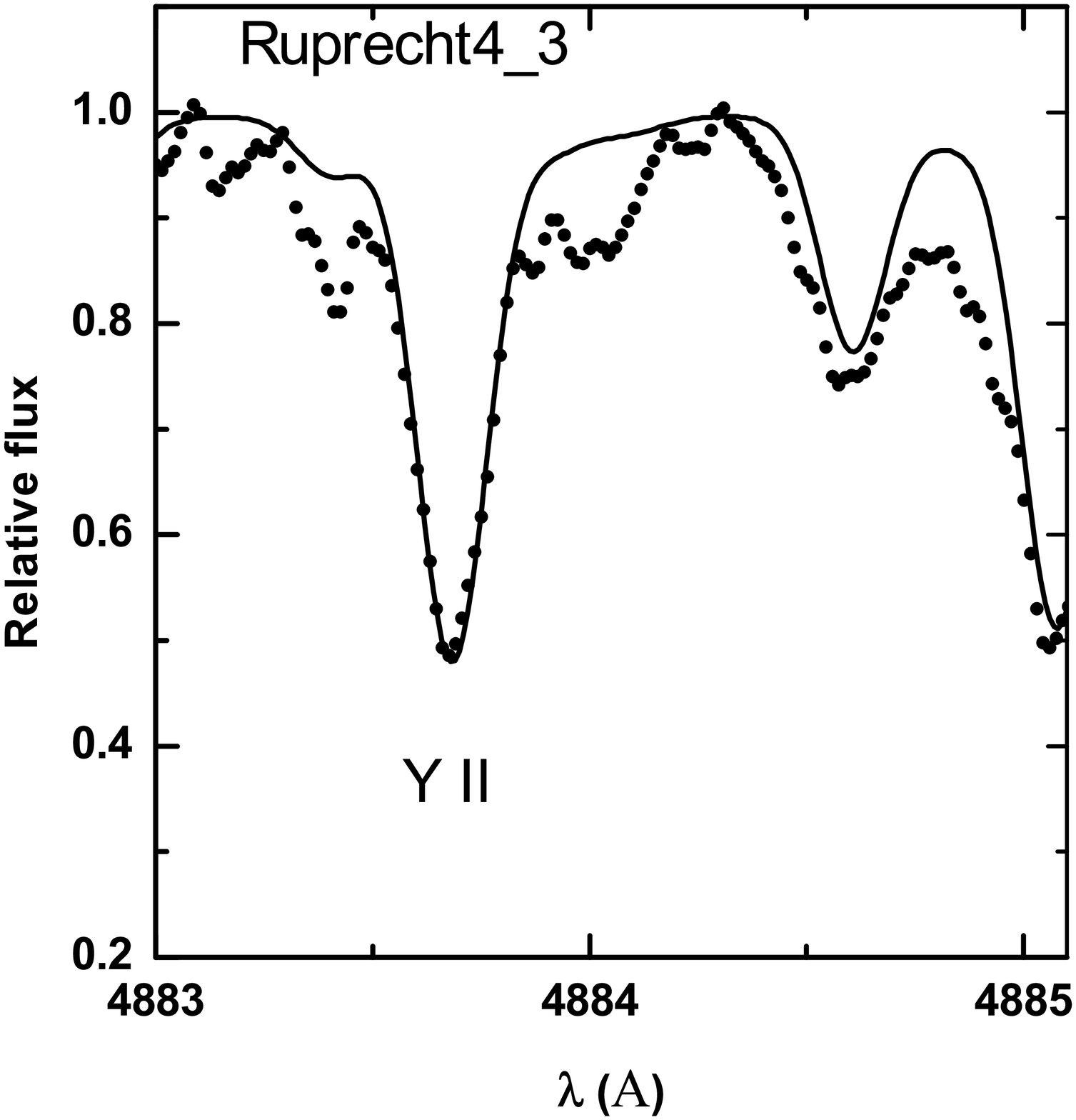}
\includegraphics{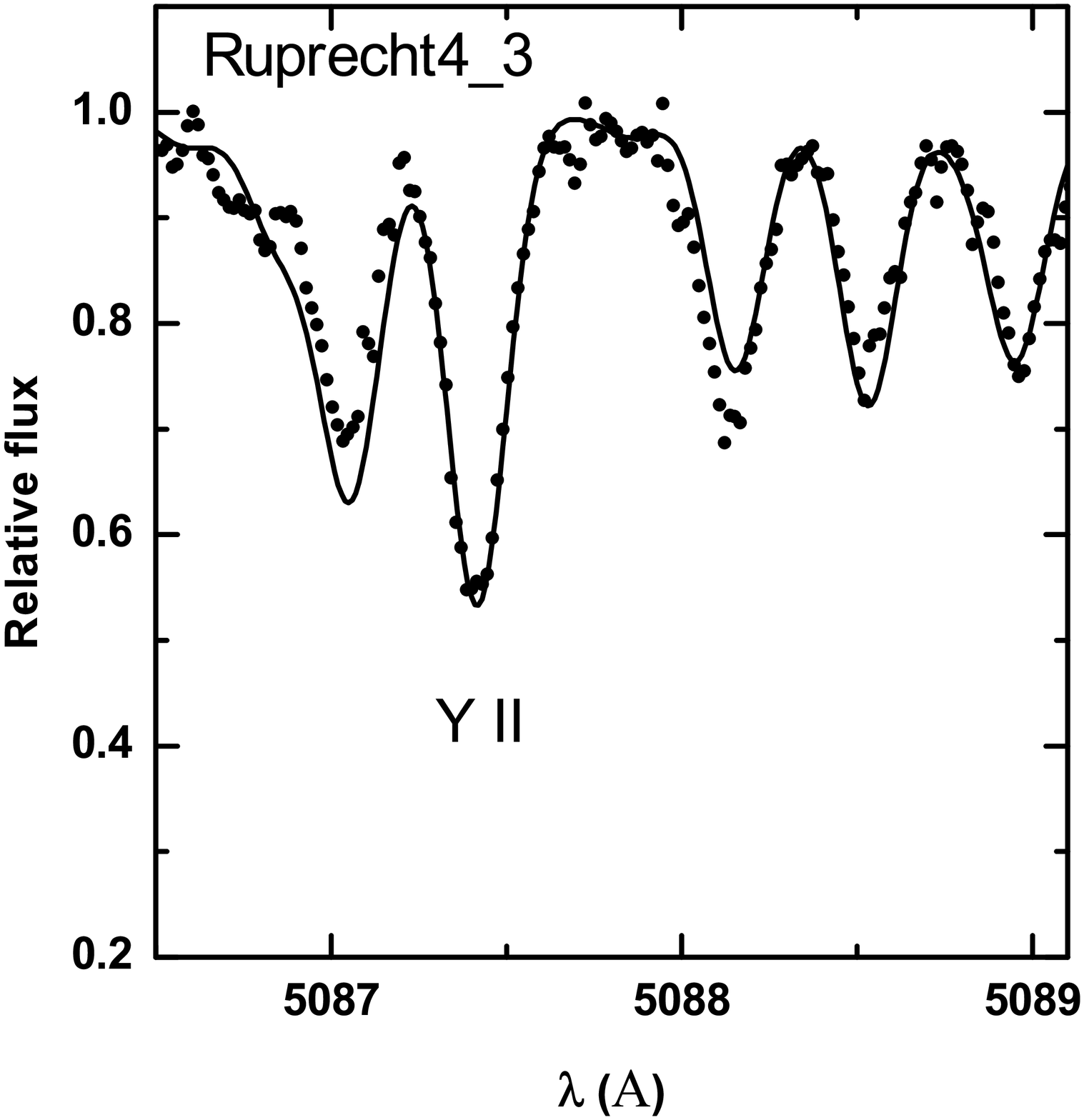}}
\caption[]{
Observed spectrum (dots) fitted by the synthetic spectrum (solid line) in the 
region of the yttrium lines Y II 4854.873, 4883.690, 
5087.426 \AA~ (the yttrium abundance (Y/H) = 2.17,  where log A(H) = 12 
for star Ruprecht~4 ID3)}
\label{yr4}
\end{figure*}

\begin{figure*}
\resizebox{\hsize}{!}
{\includegraphics{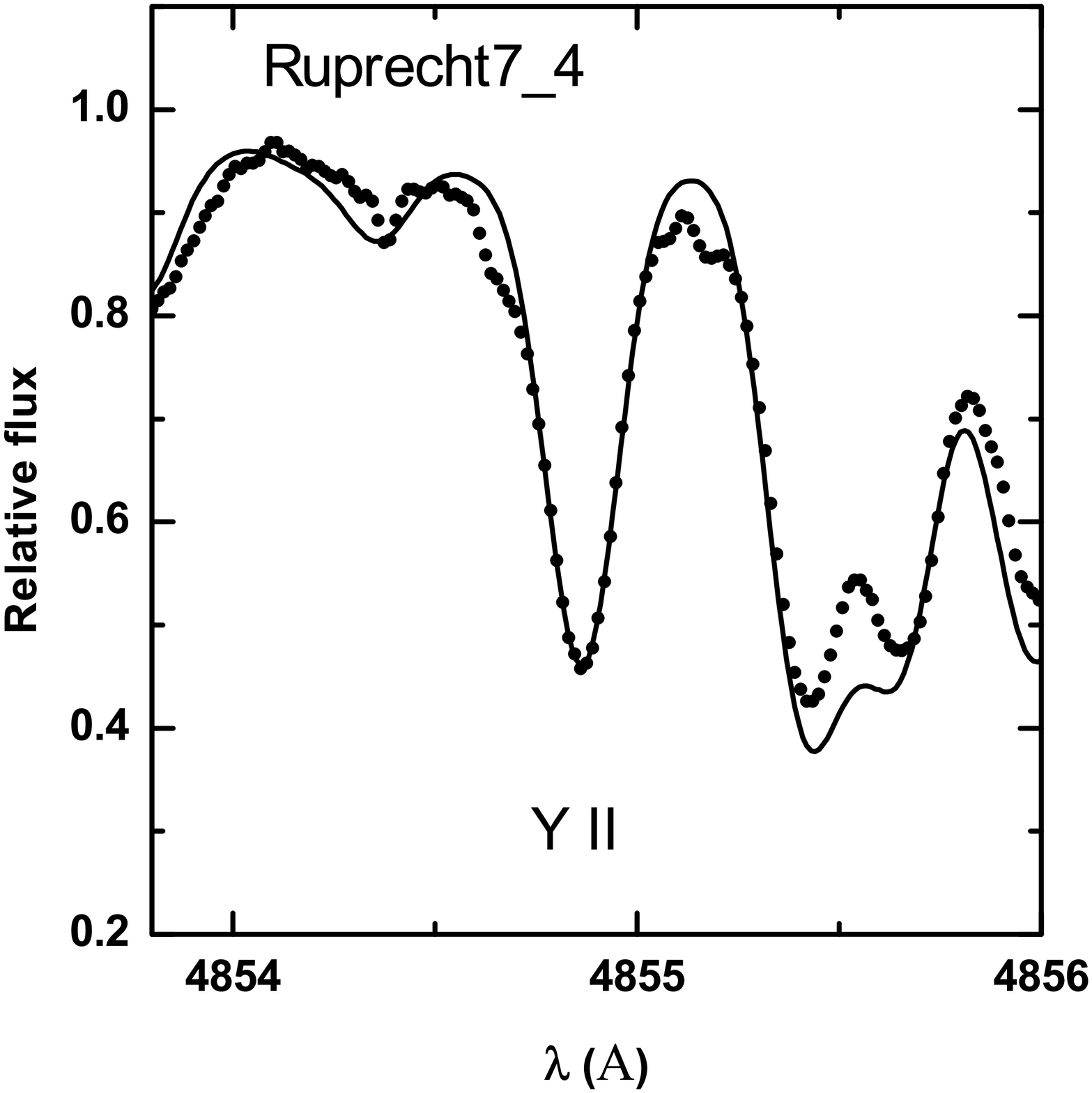}
\includegraphics{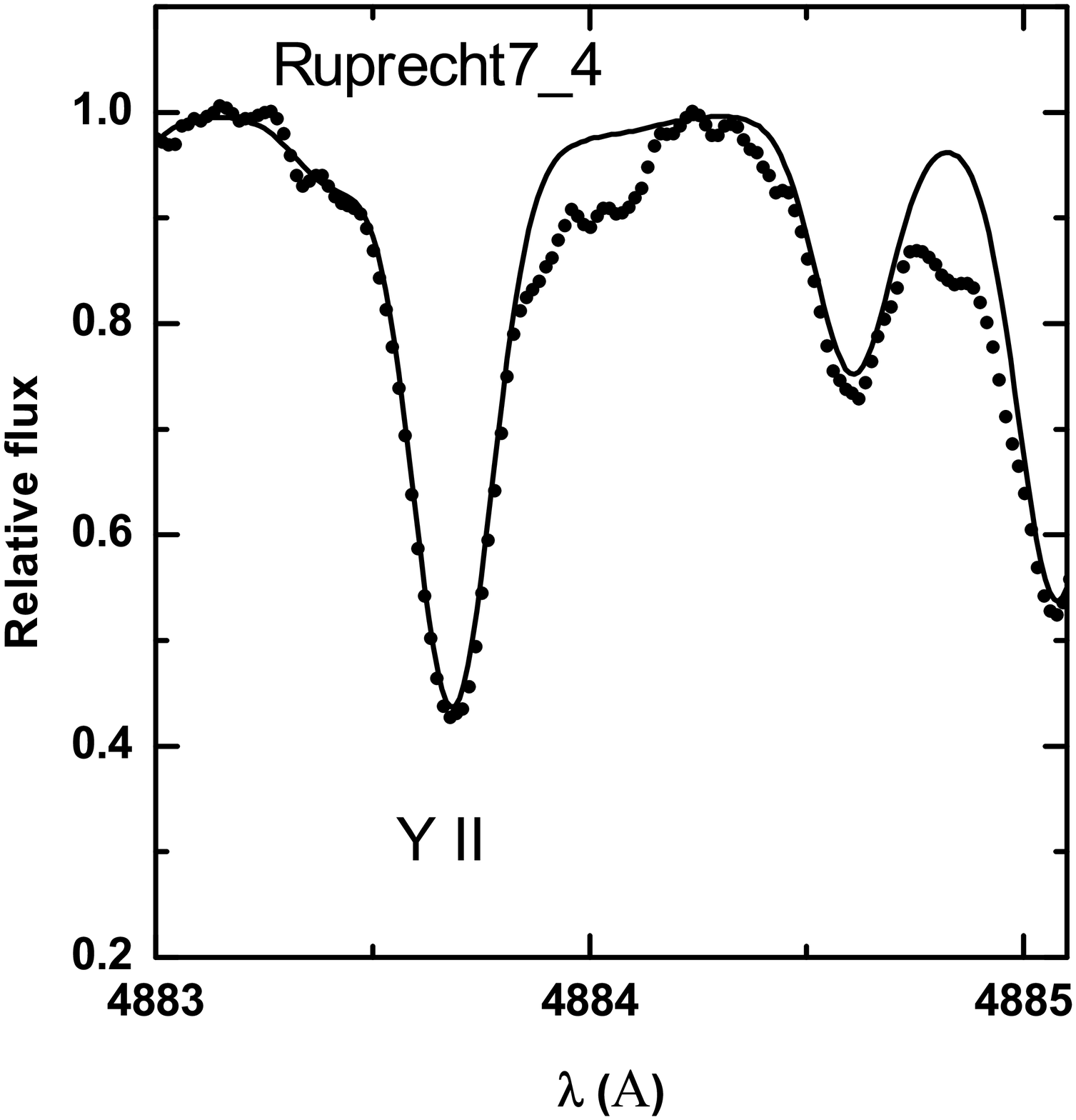}
\includegraphics{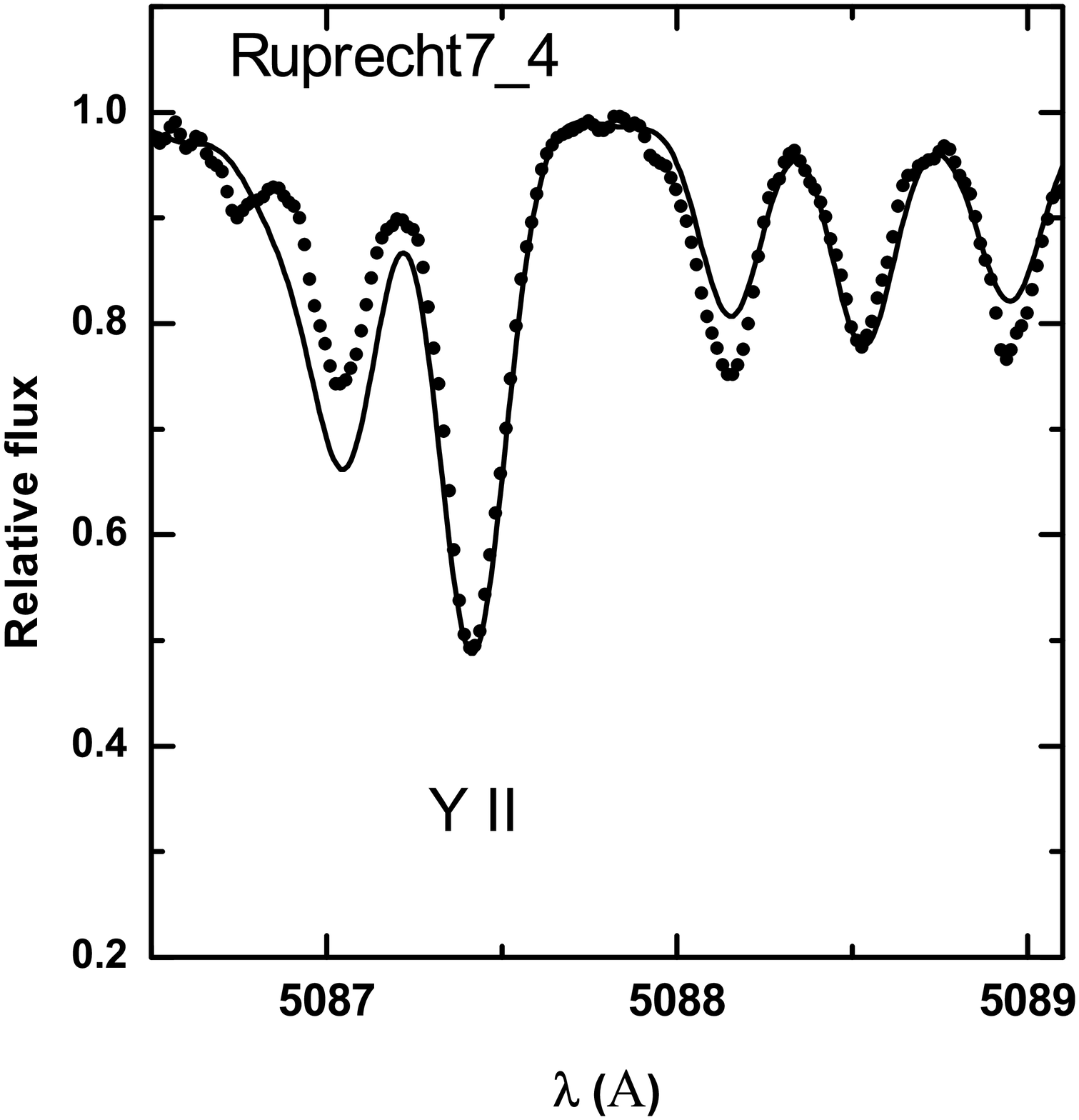}}
\caption[]{Observed spectrum 
(dots) fitted by the synthetic spectrum (solid line) in 
the region of the yttrium lines Y II 4854.873, 4883.690, 5087.426 \AA~ (the 
yttrium abundance (Y/H) = 2.05 with the hydrogen abundance = 12 for star 
Ruprecht~7 ID4).}
\label{yr7}
\end{figure*}

To estimate  abundances, we used the models by \citet{ck04}, computed for 
the atmospheric parameters of each star. The estimate of the LTE yttrium abundance 
was performed with a new version of the STARSP software package \citep{tsy96} 
and the VALD atomic data \citep{kup99} using the  three lines of Y II  centered 
at 4854.873, 4883.690, and 5087.426  \AA, which we reliably identified in 
the spectrum. The adopted Solar yttrium abundance is (Y/H)sol =  2.24 where 
log A(H)  = 12.   Examples of the synthetic spectrum fit on the observed 
spectrum in the area of the yttrium lines are shown in  Figs. \ref{yr4} and 
\ref{yr7}.

The barium abundance was estimated using the NLTE  approximation with a version 
of MULTI \citep{car86}, modified by S.A. Korotin (\citealt{an09}; 
\citealt{kor11}). Unfortunately the line 4554 \AA~ is outside the spectral
range in studied spectra and to determine the barium abundance, we used three lines of 
Ba~II (5853, 6141 and 6496 \AA) in the NLTE approximation. Our barium model 
contains 31 levels of Ba~I, 101 levels of Ba~II  with $n < 50$, and the ground 
level of Ba~III ion.
After close inspection, we included 91 bound-bound transitions as well. 
The odd barium isotopes have hyperfine splitting of their levels and, thus, 
present several Hyper Fine Structure (HFS) components for each line 
\citep{rut78}. Therefore, lines 4554~\AA~ and 6496~\AA~ were fitted by adopting 
the even-to-odd abundance ratio of 82:18 \citep{cam82}. 
It is evident that the HFS for lines 5853~\AA~ and 6141~\AA~ is not significant. 
The solar barium abundance was assumed to be (Ba/H)sol =  2.17 where  
log A(H) = 12. 
That value was obtained from the Solar Atlas \citep{kur84} with the 
same atomic data, which had been used to estimate the barium abundance in the 
stellar atmospheres. Departures from LTE do not significantly affect Ba 
abundances (correction about 0.1 dex). However we apply the NLTE corrections 
to avoid the systematic errors.

\begin{figure*}
\resizebox{\hsize}{!}
{\includegraphics{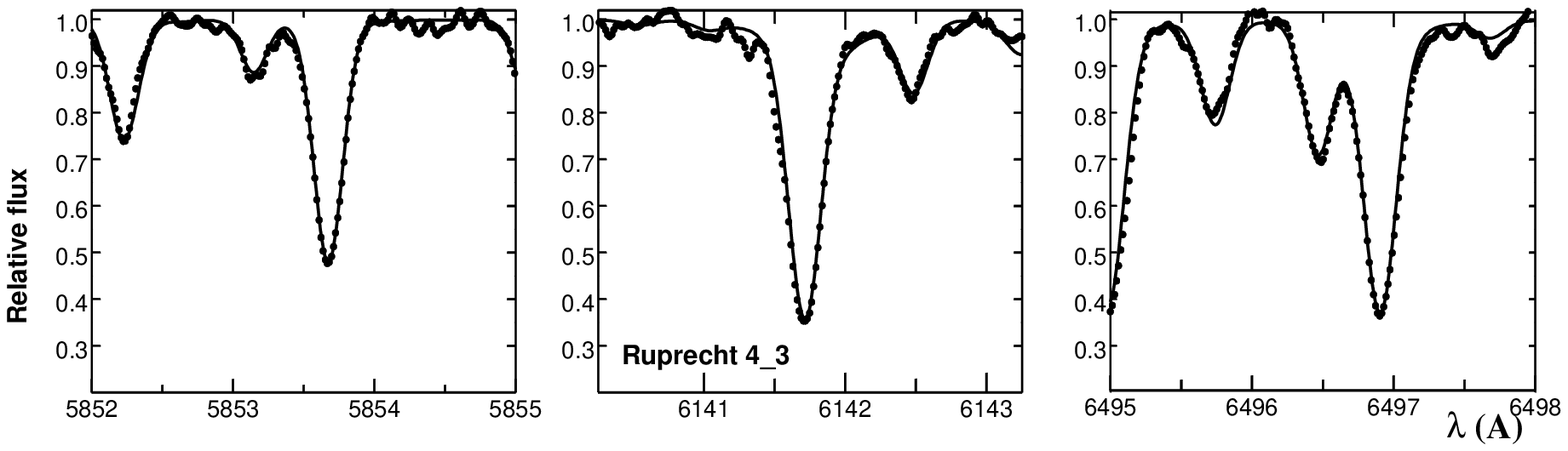}}
\resizebox{\hsize}{!}
{\includegraphics{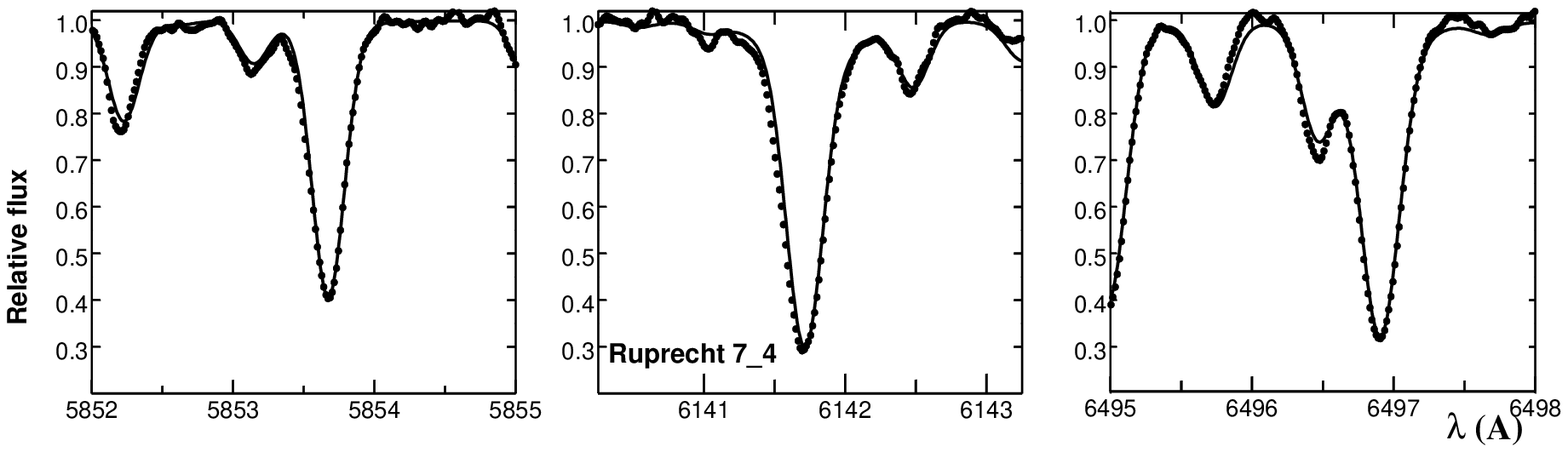}}
\caption[]{Example of the match 
of the observed spectrum  (dots) by the estimated spectrum (solid line) in the 
area of the barium lines (the barium abundance 
(Ba/H) = 2.46 is  for star Ruprecht4 ID3 and 
(Ba/H) = 2.49 is for star Ruprecht7 ID4).}
\label{bar}
\end{figure*}

Examples of the observed spectra fitted with the model spectra in the area 
of the barium lines are shown in Fig. \ref{bar}.

\begin{figure}
\resizebox{\hsize}{!}
{\includegraphics{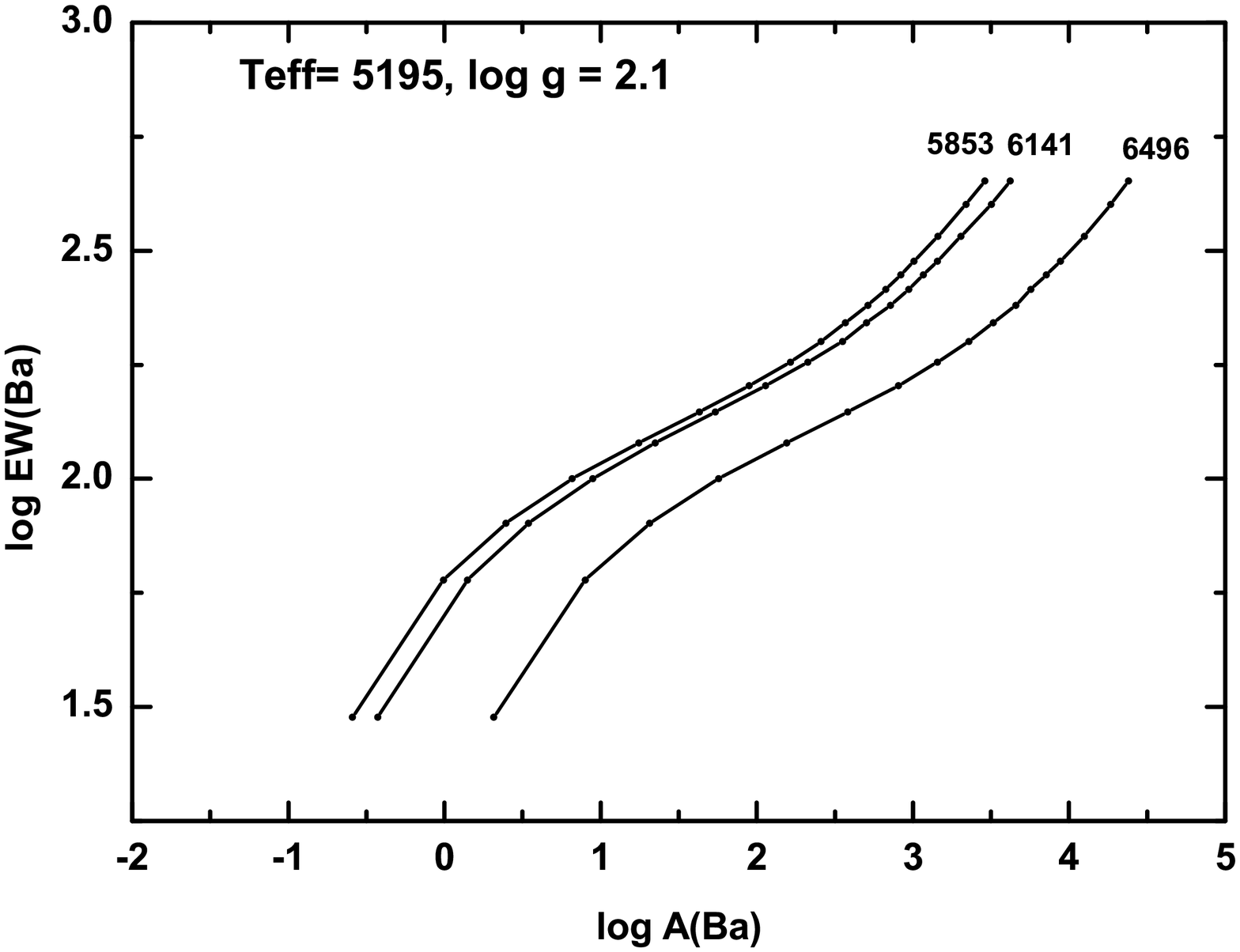}}
\caption[]{The relationship  between the equivalent width 
of the barium lines log EW(Ba) and the number of atoms, involved in the 
formation of that line log A(Ba) (for four lines of Ba II with the model of 
star Ruprecht7 ID4).}
\label{baew}
\end{figure}

\begin{figure}
\resizebox{\hsize}{!}
{\includegraphics{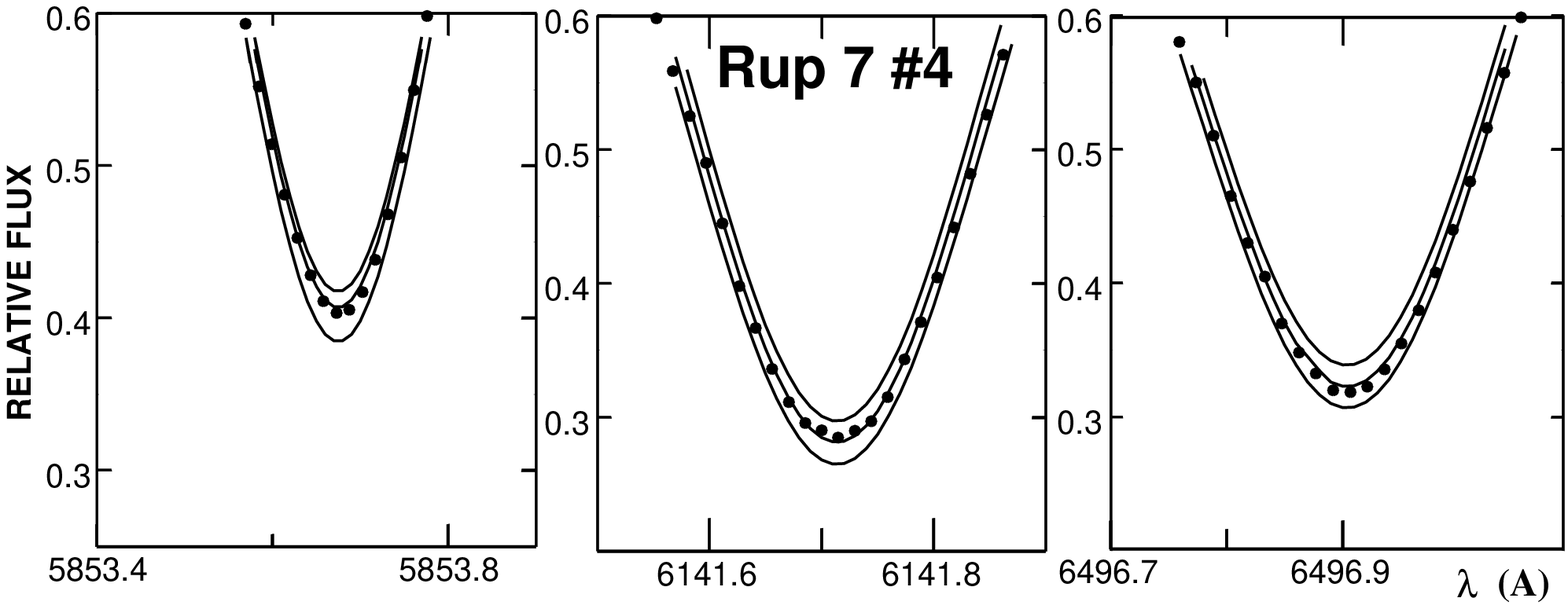}}
\caption[]{The fitting of the observed core profile of the 
barium lines and the estimated profiles with different barium 
abundances +-0.1 dex. The middle synthesis line in each panel is the "best fit".}
\label{bacor}
\end{figure}

In addition, we considered the problem associated with the saturation of the 
barium lines, since their equivalent widths are rather large. 
Fig. \ref{baew} shows the curve of growth for four barium lines, as well 
as the fitting of the observed profile of the barium line and the estimated 
profiles with different barium abundances $\pm$ 0.1 dex (Fig. \ref{bacor}). 

The estimates presented in the Figs. \ref{baew} and \ref{bacor} indicate that 
the equivalent widths and profiles are rather sensitive to the barium abundance. 
Relatively weaker and moderate lines (up to 200 m\AA~) are very sensitive to the 
element abundance; while the stronger lines, obtained by the computed synthetic 
spectra, allow one to to obtain  abundance values with an accuracy of not less 
than $\pm$ 0.1 dex (see Figs. \ref{baew} and  \ref{bacor}).

The influence of the uncertainty in  the parameters, computed by \citet{car07}, on 
the accuracy of determination of the yttrium and barium abundances 
for  Ruprecht4\_4 is given in Table \ref{ranpar}. The typical errors in 
temperature,  surface gravity logarithm and microturbulent velocity are 
$\pm$ 100 K (col 1), $\pm$ 0.2 (col 2) and $\pm$ 0.1 km/s (col 3), respectively.  
When estimating the total error (col 5), the average accuracy of the 
determination of the clusters metallicity is assumed to be 0.1 dex (col 4) and 
the accuracy the synthetic spectrum's fitting is assumed to be 0.05 dex.

\begin {table}
\caption {Compilation of random errors due to uncertainties in the atmospheric  parameters.
Ruprecht4\_4	(Teff=5150, log g=2.52, Vt=1.66)}
\label {ranpar}
\begin{tabular}{l c c c c c}
\hline
       & 1    &  2    &  3   &   4   &    5(Total error)\\
\hline
Y      & 0    & 0.12  & 0.04  & 0.1  & 0.17 \\
Ba     & 0.02 & 0.09  & 0.06  & 0.1  & 0.15 \\
\hline  
\end {tabular}  
\end {table}   

The accuracy of the fitting of the synthetic spectrum is 0.10 dex for the 
spectra with low signal-to-noise ratio; however, the total error in 
determination of the Y and Ba abundances does not exceed 0.2 dex.

\begin{figure}
\resizebox{\hsize}{!}
{\includegraphics{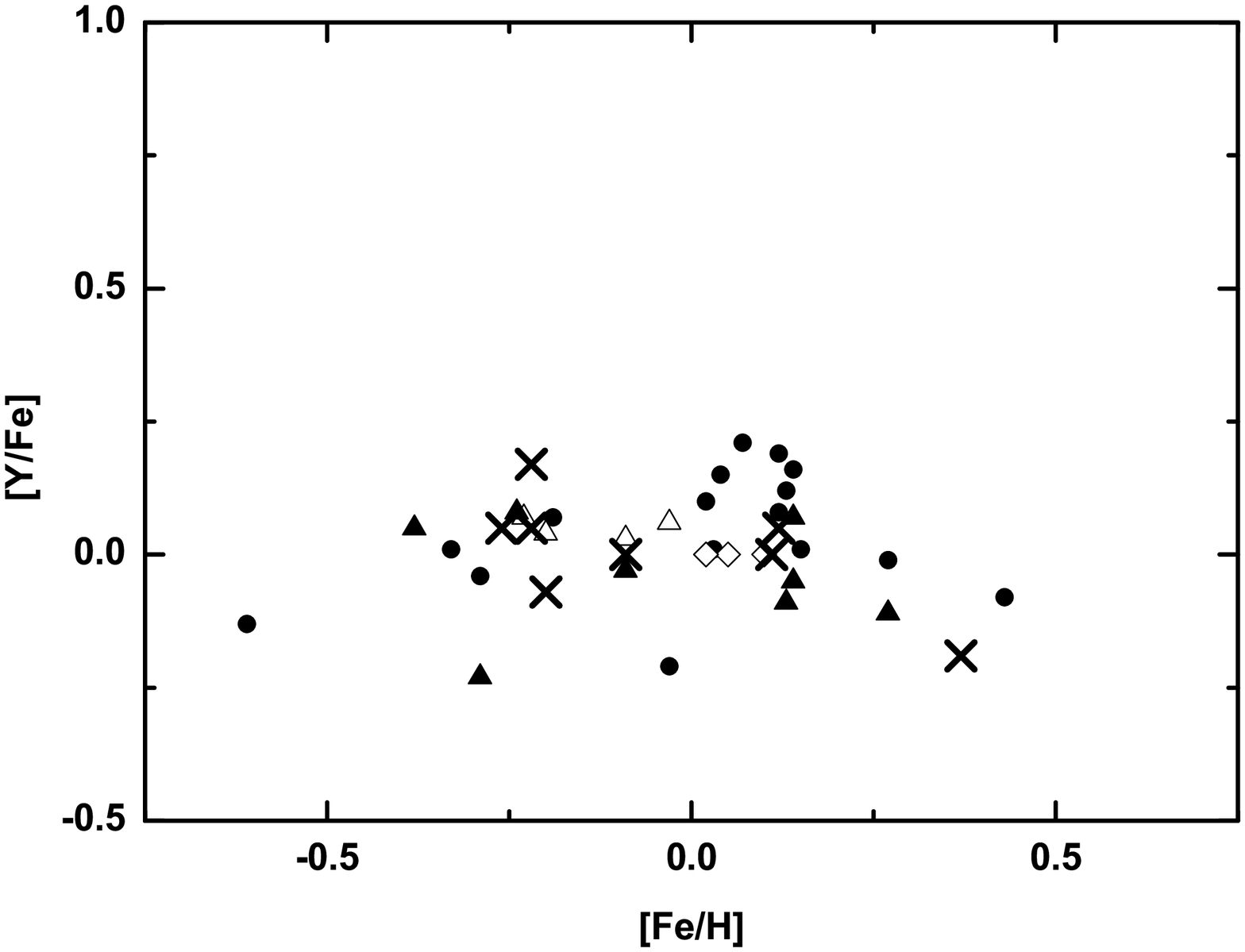}}
\resizebox{\hsize}{!}
{\includegraphics{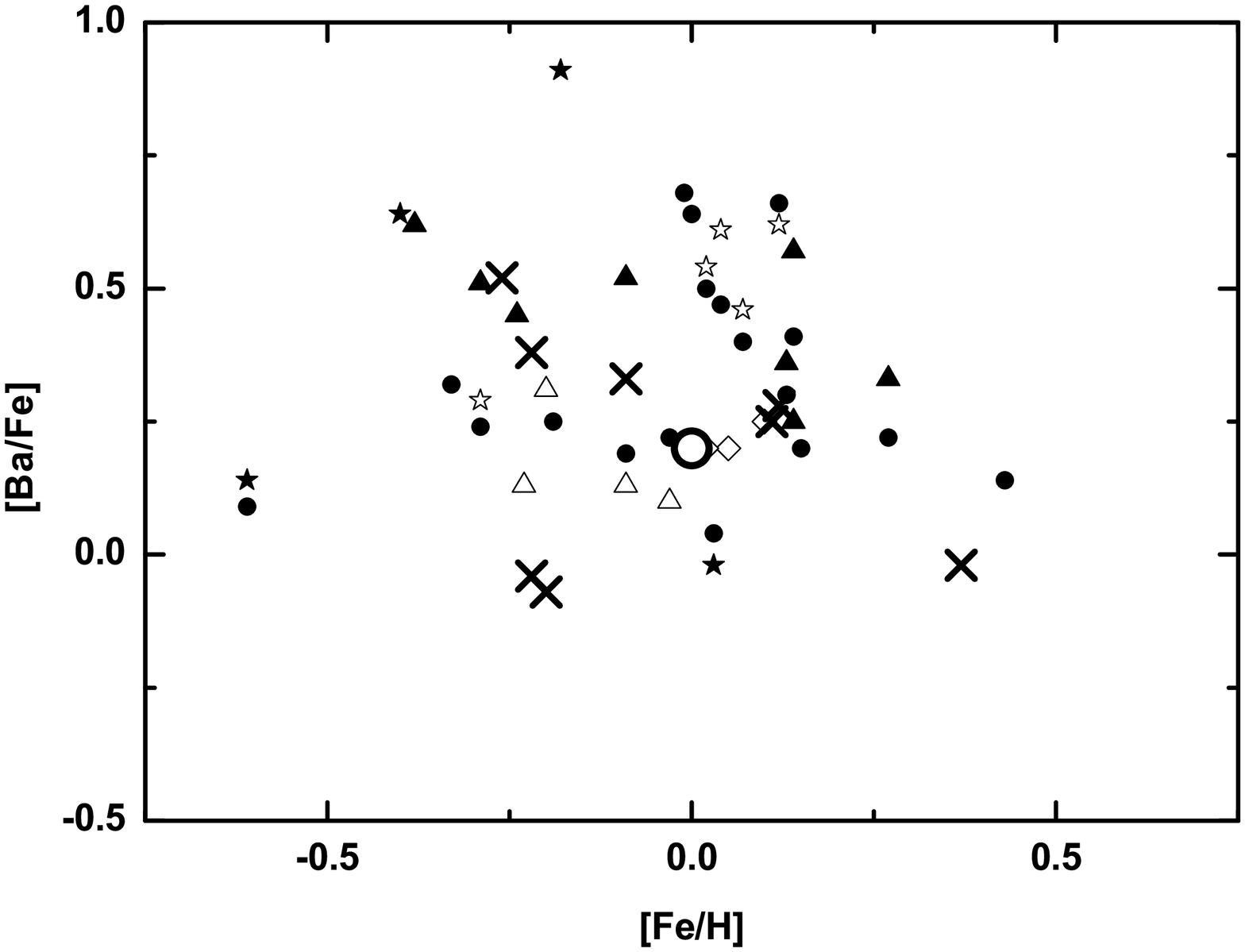}}
\caption[]{The dependence of [Y/Fe] and [Ba/Fe] on [Fe/H]. Y abundances by 
\citet{mai11} and Ba abundances by \citet{dor09} - marked as black circles; 
by \citet{pan10} and \citet{car11} - marked as black triangles; 
by \citet{dor12} - marked as open diamonds;  
by \citet{red12} - marked as open triangles; 
Ba abundances by \citet{bra08} - as open asterisks and 
by \citet{yon05} as asterisks; 
by \citet{mis12} - the thin disc (marked as black dots); 
for the Cepheids, the average values $ <$[Y/Fe]$ >$ by \citet{luc11} and  
$<$[Ba/Fe]$>$  by \citet{an13} - marked as empty circles,  
the present study - marked as crosses.}
\label{bayfe}
\end{figure}

The final results are listed in Table \ref{spar}.
The dependencies of [Y/Fe] and [Ba/Fe] on [Fe/H] for our data and those culled  
from the literaure  are presented in Fig. \ref{bayfe}.  As we can see from 
the figures,  barium shows significantly higher values and more scattering than  
yttrium.

\begin {table*}
\caption {Atmospheric  parameters and yttrium and barium abundances 
in the investigated stars.}
\label {spar}
\begin{tabular}{l c c c c c c c c c c}
\hline
Cluster	    &  ID    &  Teff & log g &  Vt   &  [Fe/H] & [Y/H] & [Y/Fe]& [Ba/H]& [Ba/Fe] &Membership\\
\hline
Berkeley 25 &   10&	5000& 2.9    &  1.65&    0.10&	-0.04&	-0.14&	-0.24&	-0.34&	NM     \\
       	    &  12&	4870&  2.75&1.5 &         -0.2&	-0.27&	-0.07&	 -0.27&	-0.07&	M      \\
	    &  13&	4860&  2.65&1.73&       -0.17&	-0.30&	-0.13 &	--&	--&	M      \\ 
Berkeley 73 &   12&	5030& 2.78& 1.4 &         -0.39&	-0.57&	-0.18&	 -0.2&	 0.19&	NM     \\
	    &  1 3&	5730& 4.15& 0.99&         0.17&	-0.1&	 0.26&	-0.12&	-0.05&	NM     \\
	    &   15&	5070& 3.12& 1.04&        -0.38&	-0.49&	-0.11&	 -0.26&	 0.12&	NM     \\
	    &   16&	4890& 2.71& 1.45&        -0.18&	-0.37&	-0.01&	 -0.2&	-0.02&	M      \\ 
	    &   18&	4940& 2.88& 1.32&        -0.27&	-0.22&	 0.05&	 -0.34&	-0.07&	M      \\ 
	    &   19&	5870& 4.23& 1.4 &         -0.03&	-0.29&	-0.26&	-0.24&	-0.21&	NM     \\ 
Berkeley 75 &    9&	4968& 2.57&1.55&          -0.44&	-0.39&	 0.05&	 -0.17&	 0.27&	NM     \\ 
	    &   22&	5180& 3.37&1.21&          -0.22&	-0.05&	 0.17&	 0.16&	 0.38&	M      \\ 
Ruprecht 4  &    3&	5180& 2.63&1.56&            0.07&	-0.07&	 0.0 &	 0.29&	 0.36&	M      \\ 
	    &    4&	5150& 2.52&1.66&           -0.04&	 0.02&	 0.06&	 0.24&          0.28&	M      \\ 
	    &    8&	5190& 2.64& 1.4 &           -0.16&	 0.03&	 0.19&	 0.18&	 0.34&	M      \\ 
	    &   18&	5040& 3.17&1.2 &            -0.35&	-0.24&	 0.09&	 -0.4&	-0.05&	NM     \\ 
	    &   29&	4920& 2.78& 1.37&          -0.34&	-0.37&	-0.03&	 -0.54&	-0.20&	NM     \\ 
Ruprecht 7  &    2&	5160& 2.12& 1.62&          -0.34&	-0.29&	 0.05&	 0.15&	 0.49&	M      \\ 
	    &    4&	5105& 2.05& 1.9 &           -0.24&	-0.2&	 0.05&	 0.32&	 0.56&	M      \\ 
	    &    5&	5230& 2.19& 2.1 &           -0.27&	-0.19&	 0.08&	 0.28&	 0.55&	M      \\ 
	    &    6&	5230& 2.23& 2.08&          -0.2&	-0.16&	 0.04&	 0.31&	 0.51&	M      \\ 
	    &    7&	5150& 2.4&1.82&             -0.25&	-0.31&	-0.06&	 0.23&	 0.48&	M      \\ 
 NGC6192    &    9&     5050& 2.3    &  1.75&  0.19 &    0.15&  -0.04&   0.43&   0.24&   M     \\  
            &   45&     5020& 2.55   &  1.60&  0.08 &    0.22&   0.14&   0.43&   0.35&   M     \\  
            &   96&     5050& 2.3    &  2.10&  0.13 &    0.10&  -0.03&   0.35&   0.22&   M     \\  
            &  137&     4670&  2.1   &  1.80&  0.07 &    0.10&   0.03&   0.4 &   0.33&   M     \\  
 NGC6404    &   5 &     5000&  1.0   &  2.60&  0.05 &    0.11&   0.06&   -- &   -- &   M     \\  
            &  16 &     4450&  1.65  &  2.10&  0.07 &    0.04&  -0.03&   0.27&   0.2 &   M     \\  
            &  27 &     4400&  1.40  &  1.80&  0.20 &    0.00&  -0.20&   0.47&   0.27&   M     \\  
            &  40 &     4250&   2.3  &  1.40&  0.11 &    0.18&   0.07&   0.39&   0.28&   M     \\  
 NGC6583    &  46 &     5100&  2.95  &  1.45&  0.40 &    0.21&  -0.19&   0.38&  -0.02&   M     \\  
            &  62 &     5050&   2.75 &  1.45&  0.34 &  --&  --&   0.27&  -0.07&   M     \\  
\hline  
\end {tabular}  
\end {table*}   

\section {The dependence of the yttrium and barium abundances on  age.}

\begin{figure}
\resizebox{\hsize}{!}
{\includegraphics{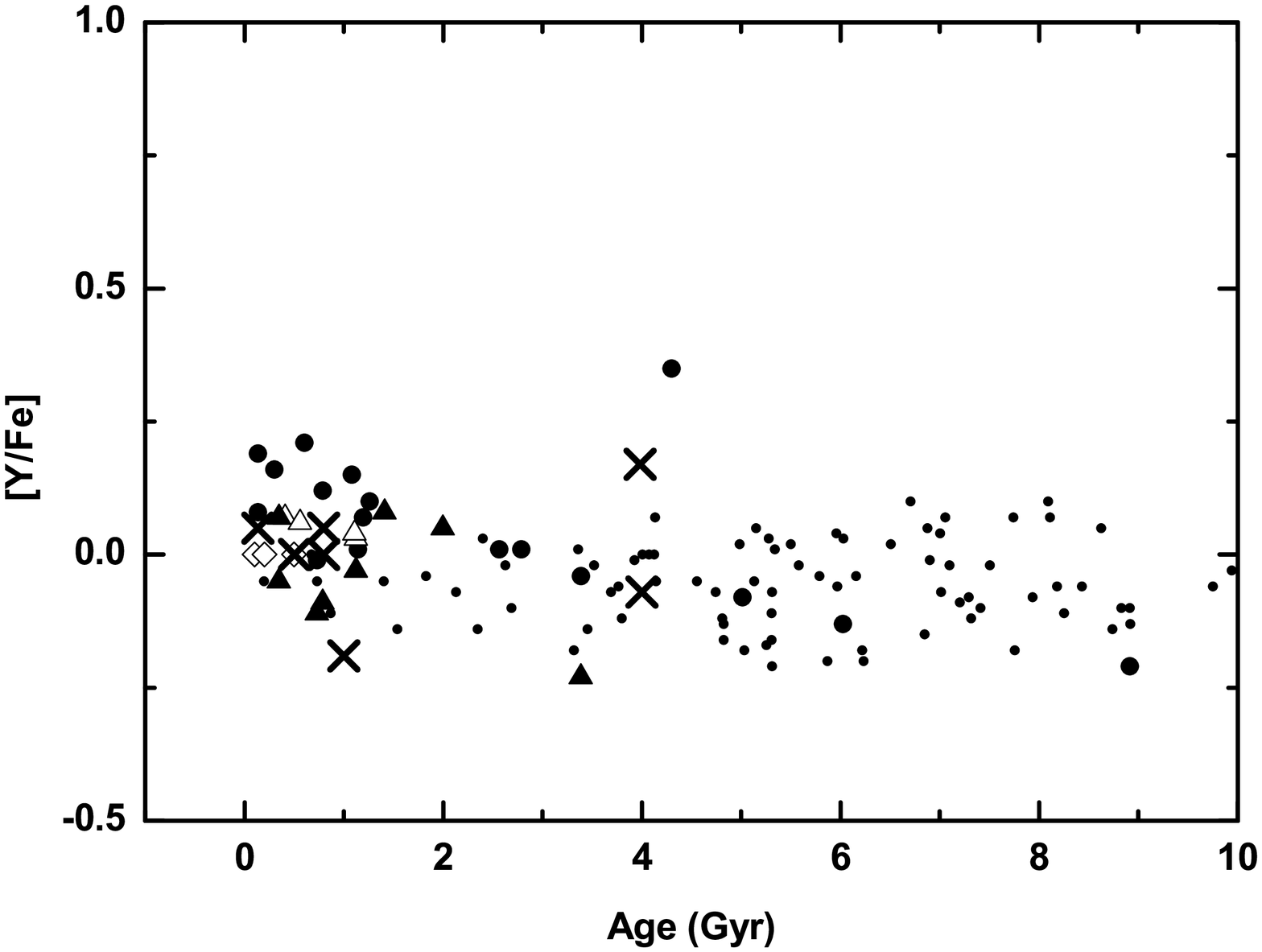}}
\resizebox{\hsize}{!}
{\includegraphics{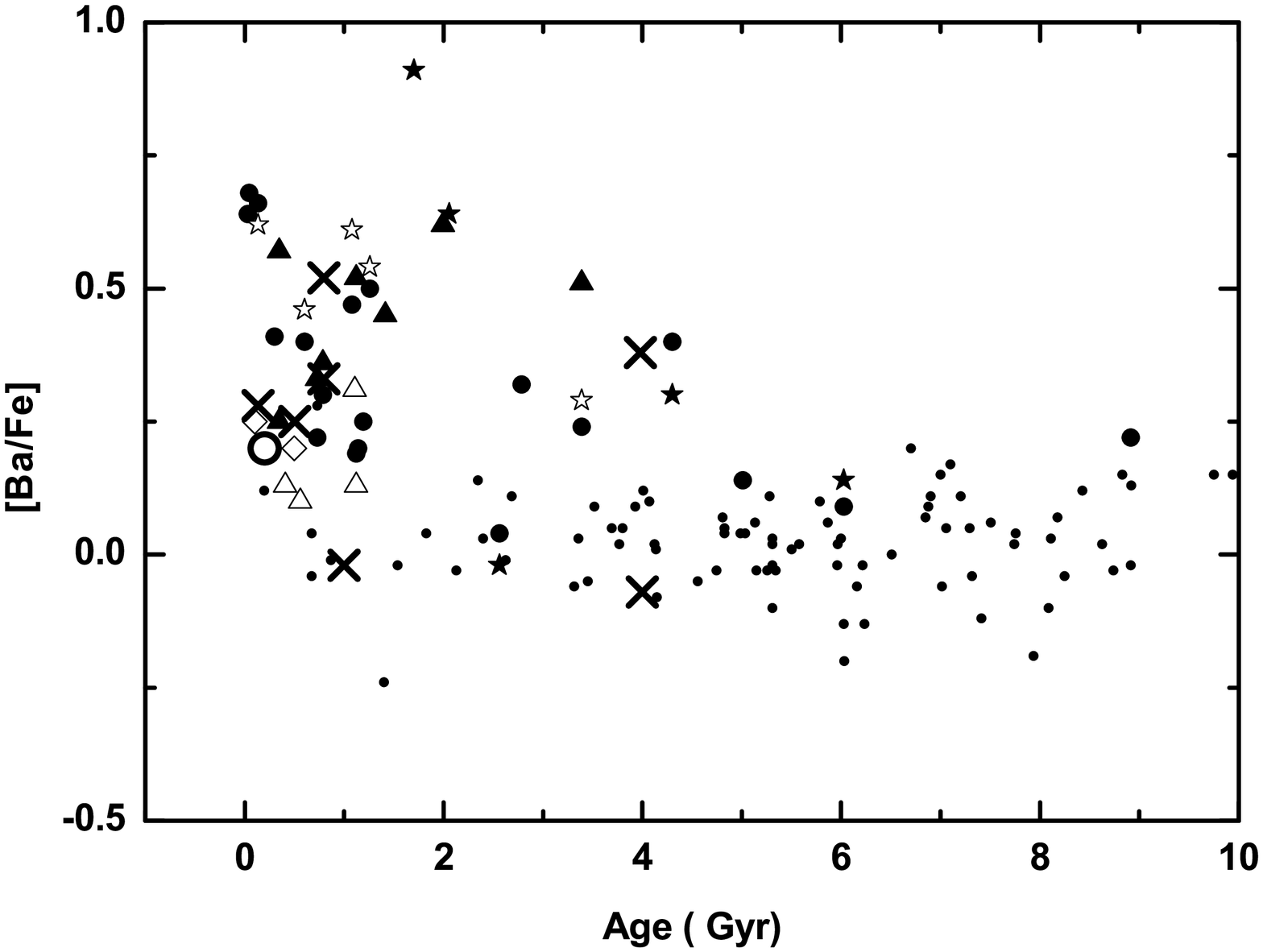}}
\caption[]{Symbols are the same as in the Fig. \ref{bayfe}.}
\label{yage}
\end{figure}

We compared (see Fig. \ref{yage})  the obtained values 
of the yttrium [Y/Fe] and barium [Ba/Fe] abundances with 
estimates  from other authors, for a number of open clusters 
(\citealt{dor09}; \citealt{pan10}; \citealt{car11}; 
\citealt{mai11}; \citealt{dor12}; \citealt{red12}; \citealt{bra08}; 
\citealt{yon05}). 
The data for the thin disc stars were taken from the study by \citet{mis12}.  
As is seen in Fig. \ref{yage}, there is a slight trend for yttrium 
(from -0.2 dex to 0.2 dex) with  increasing age. 
The data for the same young clusters obtained by \citet{car11}, differ from 
those by \citet{mai11} on average of 0.2 dex, which, nonetheless,  is within 
the range of the claimed accuracy.

Our yttrium abundances for other young clusters are also slightly higher than 
\citet{car11}; however, they nicely agree with the results 
from other authors.  Overall, the behavior of the yttrium abundance 
nicely agrees with modern chemical evolution models (with the possible 
exception of the outlier cluster Berkeley 29, age 4.3 Gyr, [Y/Fe] = 0.35 dex  
\citep{mai11}.

The fact that the correlation of the yttrium abundance and the ages of the 
stars is similar for the clusters and the thin disc favors a  
similarity of origin of the stars of the thin disc and open clusters. As  
shown by \citet{ros10}, open clusters delivered up to 40\% of the disc stars 
over the typical  Galactic disc life-time.

The moderate trend  we detect can be indicative of the increase in 
contributions into the enrichment of the interstellar medium from the 
least-massive AGB stars ($M < 1.5 M_{\sun}$).

As for barium, we observe a significant trend and a large spread of the
values. The obtained average values for the  two youngest clusters Ruprecht~4 
and Ruprecht~7 (with ages of about 0.8 Gyr) are $<$[Ba/Fe]$>$ = 0.33 and 0.56 
dex, respectively. A rather high value was also determined for cluster 
Berkeley~75 (with an age of about 3 Gyr), namely  [Ba/Fe] = 0.38 dex for 
star $\#22$. Thus, young cluster stars exhibit different 
barium abundances from moderate values of 0.2--0.3 dex to 0.6 dex.

The behavior of moderately over-abundant barium  objects is in line with the 
yttrium behavior. However, neither higher values of [Ba/Fe] nor its sharp 
increase at young ages can be explained at the moment.

In an attempt to better understand what is happening, we included in the 
discussion the yttrium and barium abundances in Galactic Cepheids (the disc 
young stars with ages around  0.2--0.4 Gyr)  as determined by 
(\citealt{luc11} (Y), \citealt{an13} (Ba)). 
In Fig. \ref{yage}, the average values are marked with large 
empty circles. A good agreement is observed for the yttrium, but the average 
values of barium in the Cepheids are significantly lower than 0.6 dex,  while 
the values of [Ba/Fe] themselves are within the range from -0.05 dex to 0.4 dex 
with the average value of 0.2 dex \citep{an13}. Taking errors into account,
the spread in barium values is the same as
in open clusters.
 
The problem of barium abundance remains, therefore, unsolved in the framework
of chemical evolution models. It is also difficult to accept the idea that all 
giants in open clusters showing such over-abundance can fall in the 
{\it Barium stars} class \citep{BK51}.

\section {The dependence  of  Y and Ba abundances on the 
Galactic location for younger  clusters}

\begin{figure}
\resizebox{\hsize}{!}
{\includegraphics[]{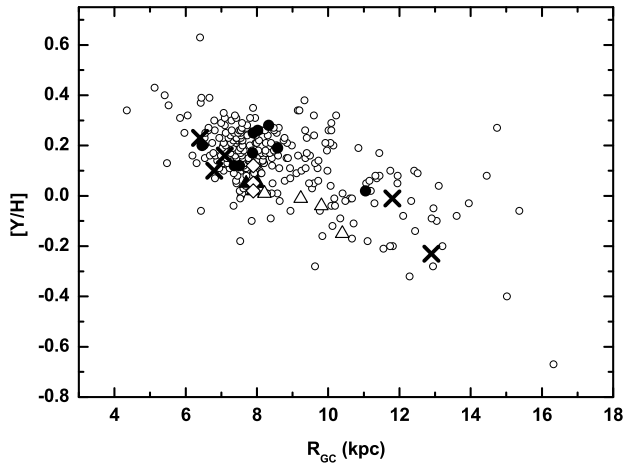}}
\caption{The dependence of the yttrium abundance on the galaxy-centric 
distance for young clusters. Symbols are the same as in Fig. \ref{bayfe}.}
\label{yhrg}
\end{figure}

\begin{figure}
\resizebox{\hsize}{!}
{\includegraphics[]{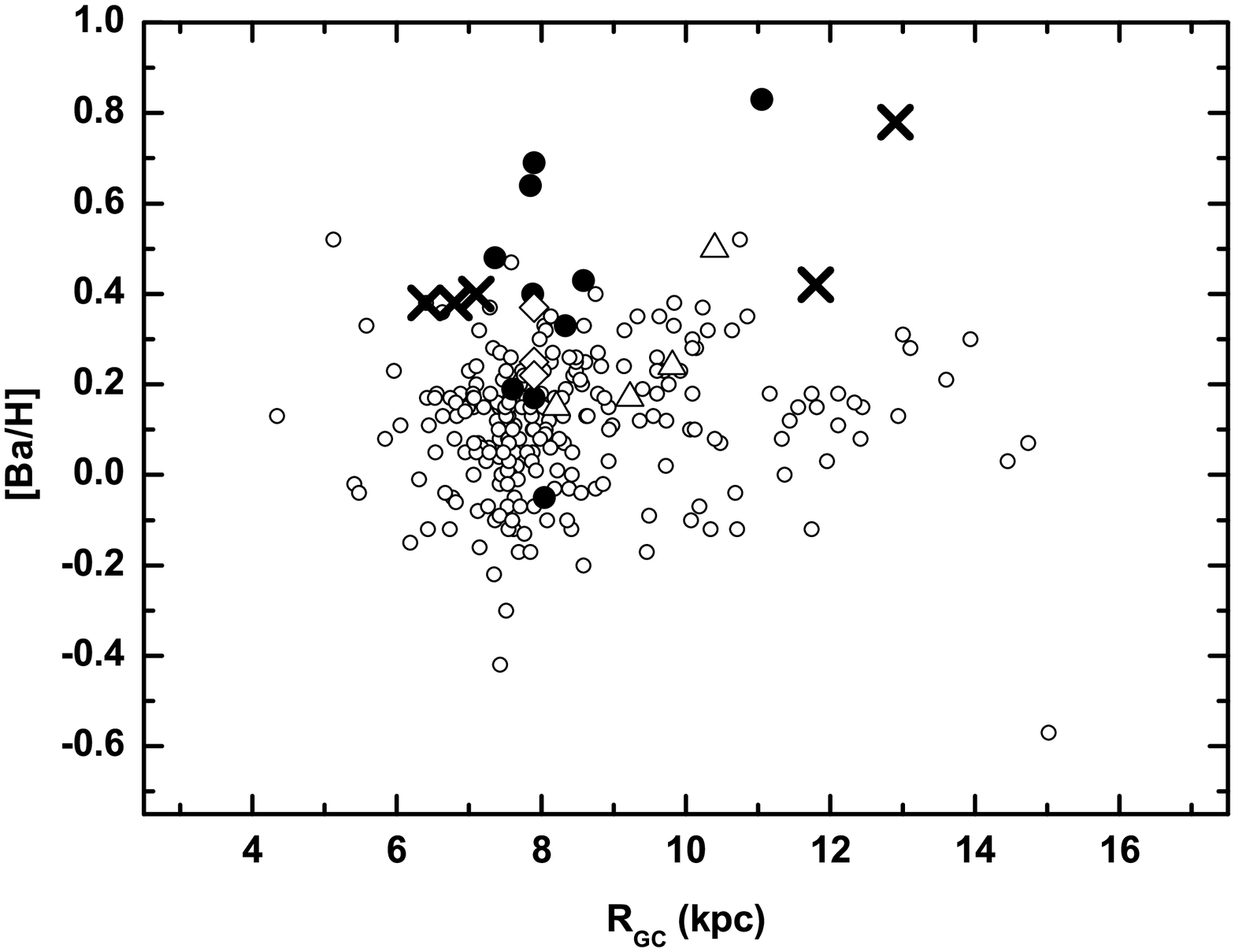}}
\caption{The dependence of the barium abundance on the galaxy-centric 
distance for young clusters. Symbols are the same as in Fig. \ref{bayfe}.}
\label{bahrg}
\end{figure}

To determine the correlation between the yttrium and barium abundances and 
the galaxy-centric distances, we compared the mean values of $<$[Y/H]$>$ and 
$<$[Ba/H]$>$ for young clusters -both from us (Ruprecht 4, Ruprecht 7, 
NGC 6192, NGC 6404, NGC 6583) and from other authors-   with those for  
Cepheids and young stars of the Galaxy disc (\citealt{luc11} (Y); 
\citealt{an13} (Ba)) (see Figs. \ref{yhrg} and \ref{bahrg}). 
To consistently compare the data 
with those for the Cepheids, the galacto-centric distances for all clusters 
were recomputed 
adopting 7.9 kpc as Sun distance from the centre of the Galaxy. 
The galaxy-centric distances for the program 
clusters were computed with the following formula (the distances are given 
in pc):   
\begin{equation}
{\rm  R_{\rm G} = \left[R_{\rm G,\odot}^{2} + (d\cos b)^{2} -
        2R_{G,\odot} d \cos b \cos l\right]^{1/2}}
\end{equation}
where R$_{\rm G,\odot}$ is the galacto-centric distance of the Sun, d is the 
heliocentric distance of a cluster, l is the galactic longitude, and b is the 
galactic latitude. The galacto-centric distance of the 
Sun R$_{\rm G,\odot}$ = 7.9 kpc was adopted from the recent determination 
by  \citet{mcnet00}. 
The distribution of the analyzed clusters in the galactic plane is shown in 
Figs. \ref{yhrg} and \ref{bahrg} .

The yttrium abundance in young open clusters mirrors the trend of Cepheids and 
confirms the existence of an yttrium abundance gradient in the Galactic disc.

Barium behaves differently. 
The Ba abundances in a number of clusters are within the same area as the 
Ba values for the Cepheids, but four clusters exhibit significant barium 
over-abundances that exceed the spread of values ([Ba/H] $>$ 0.5 dex) in the 
Cepheids. Those are  IC 2391, NGC 2324, IC 2602 \citep{dor09} and Ruprecht 7 
(our determinations). 
The existence of such a sizable overabundance, especially for the 
Ruprecht~7, remains an open issue. 
This cannot be imputed to deviations from the LTE, since  our estimates were 
obtained taking these effects into proper account. Moreover,  as  shown by 
\citet{dor12}, this cannot be related to chromospheric activity.

We note that Ruprecht~7, besides showing a net over-abundance, also possesses
an abnormal orbit according to the calculations by \citet{vp10}, 
\citet{goz12a}. Unfortunately, however,  the accuracy 
in the orbit determination (large proper motion errors for distant clusters, 
and the adopted fixed Galacti potential) does not allow us to further explore 
the possible consequences of this result.

Other scenarios have to be explored. We speculate that they possibly formed 
out of  a metal poor environment previously contaminated by the close passage 
of high velocity clouds, or old, metal poor systems like globular clusters or 
dwarf galaxies.

\section{Summary and conclusions}

In this paper, yttrium and barium abundances are determined for 30 stars, 
including 22 stars in eight open clusters using modern techniques of chemical 
abundance analysis.

Our main conclusions can be summarized as follows:

\begin{description}

\item $\bullet$  The correlation between the yttrium abundance and the age is 
similar for the 
clusters and the thin disc, which seems to indicate similarity of origin for
the Galactic thin disc and open clusters. 
The moderated yttrium over-abundance in young systems
can be indicative of an increase of the contribution of 
 the least-massive AGB stars with $M < 1.5 M_{\sun}$. 
The trend of yttrium for young open clusters with the galacto-centric distance 
is similar to the one obtained for the Cepheids, and it confirms the 
existence of an  yttrium abundance gradient in the Galactic disc. 

\item $\bullet$  This sample of clusters shows significant barium overabundance, 
including young clusters as Ruprecht~7. 
Based on these results,  we can suggest that barium 
overabundance is real and is not due to errors in the abundance determination.
As an alternative,  one has to invoke that this family of  clusters is somehow 
peculiar, or their ambient formation has been contaminated in a very special way.

\end{description}

We conclude by suggesting that the study of the barium 
abundance (including  the hyperfine structure and in the approximation 
of the NLTE) in as many stars and clusters as possible (especially young ones) 
is very welcome, and additional data should be provided.

\section*{Acknowledgments}
 The authors thank the referee Chris Sneden for carefully reading the article and useful suggestions that improved the presentation of the results. We express our gratitude to Sandy Strunk for reading carefully this manuscript.
T.M. and S.K. thank for the support from the Swiss National Science Foundation, 
project SCOPES No. IZ73Z0-128180.

\label{lastpage}

\bsp


\begin{thebibliography}{}

\bibitem[\protect\citeauthoryear{Alonso et al.}{1999}]{AAMR99}
Alonso A., Arribas S., Mart\'inez-Roger C., 1999, A\&AS, 140, 261

\bibitem[\protect\citeauthoryear{Andrievsky et al.}{2009}]{an09}  
Andrievsky S.M., Spite M., Korotin S.A. et al., 2009, A\&A, 494, 1083

\bibitem[\protect\citeauthoryear{Andrievsky et al.}{2013}]{an13}  
Andrievsky S.M.,  L\'epine J.R.D, Korotin S.A., Luck R.E., 
Kovtyukh V.V., Maciel W.J., 2013, MNRAS, 428, 3252

\bibitem[\protect\citeauthoryear{Bidelman \& Keenan}{1951}]{BK51} Bidelman
W.P., Keenan P.C., 1951, ApJ, 114, 473

\bibitem[\protect\citeauthoryear{Bragaglia et al.}{2008}]{bra08}
Bragaglia A., Sestito P., Villanova S., Carretta E., Randich S., Tosi M., 
2008, A\&A, 480, 79

\bibitem[\protect\citeauthoryear{Brewer \& Carney}{2006}]{BC06}
Brewer M.-M., Carney B.W., 2006, AJ, 131, 431

\bibitem[\protect\citeauthoryear{Burbidge et al.}{1957}]{BBFH57}
Burbidge E. M., Burbidge G. R., Fowler W.A., Hoyle F., 1957, 
Reviews of Modern Physics, 29, 547

\bibitem[\protect\citeauthoryear{Busso et al.}{1999}]{bu99}  
Busso M., Gallino R., Wasserburg G. J., 1999, ARA\&A 37, 239

\bibitem[\protect\citeauthoryear{Busso et al.}{2001}]{bu01}  
Busso V., Gallino R., Lambert D., Travaglio C.,  Smith, V.,
2001, ApJ, 557, 802

\bibitem[\protect\citeauthoryear{Cameron}{1982}]{cam82} 
Cameron A.G.W., 1982, Ap\&SS, 82, 123

\bibitem[\protect\citeauthoryear{Carlsson}{1986}]{car86} 
Carlsson M., 1986, Uppsala Obs. Rep., 33

\bibitem[\protect\citeauthoryear{Carraro et al.}{2005a}]{car05a}
Carraro G., Geisler D., Moitinho A., Baume G., V\'azquez R. A., 
2005a, A\&A, 442, 917

\bibitem[\protect\citeauthoryear{Carraro et al.}{2005b}]{car05b}
Carraro G., Geisler D., Baume G., V'azquez R. A., Moitinho A., 
2005b, MNRAS, 360, 655

\bibitem[\protect\citeauthoryear{Carraro et al.}{2005c}]{car05c}
Carraro G., Mendez R.A., Costa E., 2005c, MNRAS, 360, 655

\bibitem[\protect\citeauthoryear{Carraro et al.}{2007}]{car07}
Carraro G.,  Geisler D., Villanova S., Frinchaboy P. M., Majewski S. R.,
2007, A\&A 476, 217

\bibitem[\protect\citeauthoryear{Carrera \& Pancino}{2011}]{car11}
Carrera R., Pancino E.,  2011, A\&A 535, 30

\bibitem[\protect\citeauthoryear{Castelli \& Kurucz}{2004}]{ck04} 
Castelli F., Kurucz R.L., 2004, arXiv:astro-ph/0405087

\bibitem[\protect\citeauthoryear{Chiappini et al.}{1997}]{chi97} 
Chiappini C., Matteucci F., Gratton R., 1997, ApJ, 477, 765

\bibitem[\protect\citeauthoryear{Claria et al.}{2006}]{cla06} 
Claria J. J., Mermilliod J.-C., Piatti A. E., Parisi M. C.,
2006, A\&A, 453, 91

\bibitem[\protect\citeauthoryear{Cristallo et al.}{2009}]{cri09} 
Cristallo S., Straniero O., Gallino R., Piersanti L., Dominguez I., 
Lederer M.T., 2009, Ap.J., 696, 797.

\bibitem[\protect\citeauthoryear{Desidera et al.}{2011}]{des11} 
Desidera S., Covino E., Messina S., D'\,Orazi V., Alcala J. M., Brugaletta E.,
Carson J., Lanzafame A. C., Launhardt R.l., 2011, A\&A, 529, 54 

\bibitem[\protect\citeauthoryear{D'\,Orazi et al.}{2009}]{dor09}
D'\,Orazi V., Magrini L., Randich S., Galli D.,  Busso M., Sestito P., 
2009, ApJ, 693, 31

\bibitem[\protect\citeauthoryear{D'\,Orazi et al.}{2012}]{dor12}
D'\,Orazi V. , Biazzo K., Desidera S., Covino E., Andrievsky S. M., 
Gratton R. G., 2012, MNRAS, 423, 2789 

\bibitem[\protect\citeauthoryear{Galazutdinov}{1992}]{gal92} 
Galazutdinov G. A., 1992, Preprint SAO RAS., 92, 96  

\bibitem[\protect\citeauthoryear{Gallino et al.}{1998}]{gal98} 
Gallino R., Arlandini C., Busso M., et al., 1998, ApJ, 497, 388

\bibitem[\protect\citeauthoryear{Gozha, Borkova \& Marsakov}{2012}]{goz12a} 
Gozha M.L., Borkova T. V., Marsakov V.A., 2012, Astron. Letters, 38, 506


\bibitem[\protect\citeauthoryear{Jacobson, Friel \& Pilachowski}{2011}]{jfp11} 
Jacobson H. R., Friel E. D., Pilachowski C. A., 2011, AJ, 141, 58
 
\bibitem[\protect\citeauthoryear{Kappeler et al.}{1989}]{kap89} 
Kappeler F., Beer H., Wisshak K., 1989, Reports on Progress in Physics, 52, 945

\bibitem[\protect\citeauthoryear{Korotin et al.}{2011}]{kor11} 
Korotin S.A.,  Mishenina T.,  Gorbaneva T.,  Soubiran C., 
2011, MNRAS, 415, 2093

\bibitem[\protect\citeauthoryear{Kovtyukh et al.}{2003}]{kov03} 
Kovtyukh V.V., Soubiran C., Belik S.I., Gorlova N.I., 2003, A\&A 411, 559

\bibitem[\protect\citeauthoryear{Kupka et al.}{1999}]{kup99}
Kupka F., Piskunov N.E., Ryabchikova T.A., Stempels H. C., Weiss W. W.,
1999, A\&AS, 138, 119

\bibitem[\protect\citeauthoryear{Kurucz et al.}{1984}]{kur84}
Kurucz R.L., Furenlid I., Brault J., Testerman L., 1984, Solar Flux
atlas from 296 to 1300 nm, Nat. Solar Obs., Sunspot, New Mexico.

\bibitem[\protect\citeauthoryear{Kurucz}{1993}]{kur93}
Kurucz R.L., 1993, CD ROM n13

\bibitem[\protect\citeauthoryear{Luck \& Lambert}{2011}]{luc11}
Luck R. E., Lambert D. L., 2011, AJ, 142, 136 

\bibitem[\protect\citeauthoryear{Maiorca et al.}{2011}]{mai11}
Maiorca E., Randich S., Busso M., Magrini L., Palmerini S.E.,
2011, ApJ, 736, 120

\bibitem[\protect\citeauthoryear{Magrini et al.}{2010}]{mag10}
Magrini L., Randich S., Zoccali M. et al.,
2010, A\&A, 523, id.A11

\bibitem[\protect\citeauthoryear{McNamara et al.}{2000}]{mcnet00}
McNamara D. H., Madsen J. B., Barnes J., Ericksen, B. F., 2000, PASP, 112, 202 

\bibitem[\protect\citeauthoryear{Mishenina et al.}{2012}]{mis12}
Mishenina T. V., Soubiran C., Korotin S.A., Gorbaneva T.I., Basak N.Yu.,
2012, in Reyle C., Robin A., Schultheis M., eds., 
EPJ Web Conf., 19, id.05006

\bibitem[\protect\citeauthoryear{Pancino et al.}{2010}]{pan10}
Pancino E., Carrera R., Rossetti E., Gallart C., 2010, A\&A, 511, id.A56.

\bibitem[\protect\citeauthoryear{Reddy, Giridhar \& Lambert}{2012}]{red12} 
Reddy A.B.S., Giridhar S., Lambert D. L., 2012, MNRAS, 419, 1350

\bibitem[\protect\citeauthoryear{Roser et al.}{2010}]{ros10}
Roser S.R., Kharchenko N.V., Piskunov A.E. Schilbach E., Scholz R.-D., 
Zinnecker H., 2010, Astron. Nachr., 331, 519

\bibitem[\protect\citeauthoryear{Rutten}{1978}]{rut78} 
Rutten R.J., 1978, SoPh 56, 237

\bibitem[\protect\citeauthoryear{Serminato et al.}{2009}]{ser09} 
Serminato A., Gallino R., Travaglio C., Bisterzo S., Straniero O.,
2009, Publ. Astron. Soc. of Australia, 26, 153 

\bibitem[\protect\citeauthoryear{Soubiran \& Girard}{2005}]{sob05} 
Soubiran C., Girard P., 2005, A\&A, 438, 139

\bibitem[\protect\citeauthoryear{Sneden}{1973}]{sn73} 
Sneden C., 1973, ApJ, 184, 839

\bibitem[\protect\citeauthoryear{Straniero, Cristallo \& Gallino}{2009}]{str09} 
Straniero O., Cristallo S., Gallino R., 2009, Publ. Astron. Soc. of Australia,
26, 133

\bibitem[\protect\citeauthoryear{Travaglio et al.}{1999}]{tra99} 
Travaglio C., Galli D., Gallino R., Busso M., Ferrini F., Straniero O., 
1999, ApJ, 521, 691

\bibitem[\protect\citeauthoryear{Tsymbal}{1996}]{tsy96} 
Tsymbal V.V., 1996, ASP Conf. Ser., 108, 198

\bibitem[\protect\citeauthoryear{Vande Putte et al.}{2010}]{vp10} 
Vande Putte D., Garnier T.P., Ferreras I., Mignani R. P., Cropper M., 
2010, MNRAS, 407, 2109 

\bibitem[\protect\citeauthoryear{Yong et al.}{2005}]{yon05}
Yong D., Carney B. W., Teixera de Almeida M.-L., 2005, AJ, 130, 597

\end{thebibliography}
\end{document}